\documentclass{pasj01}

\RequirePackage{mathpazo}
\RequirePackage[T1]{fontenc}

\newcounter{author}
\def\authorcount#1#2{\refstepcounter{author}\label{#1}
                     \altaffiltext{\ref{#1}}{#2}}

\begin{document}
\SetRunningHead{T. Kato, F.-J. Hambsch, L. Cook}{Supercycle in V4140 Sgr}

\Received{201X/XX/XX}
\Accepted{201X/XX/XX}

\title{Detection of the Supercycle in V4140 Sagittarii:
       First Eclipsing ER Ursae Majoris-like Object}

\author{Taichi~\textsc{Kato},\altaffilmark{\ref{affil:Kyoto}*}
        Franz-Josef~\textsc{Hambsch},\altaffilmark{\ref{affil:GEOS}}$^,$\altaffilmark{\ref{affil:BAV}}$^,$\altaffilmark{\ref{affil:Hambsch}}
        Lewis~M.~\textsc{Cook},\altaffilmark{\ref{affil:LewCook}}
}

\authorcount{affil:Kyoto}{
     Department of Astronomy, Kyoto University, Kyoto 606-8502, Japan}
\email{$^*$tkato@kusastro.kyoto-u.ac.jp}

\authorcount{affil:GEOS}{
     Groupe Europ\'een d'Observations Stellaires (GEOS),
     23 Parc de Levesville, 28300 Bailleau l'Ev\^eque, France}

\authorcount{affil:BAV}{
     Bundesdeutsche Arbeitsgemeinschaft f\"ur Ver\"anderliche Sterne
     (BAV), Munsterdamm 90, 12169 Berlin, Germany}

\authorcount{affil:Hambsch}{
     Vereniging Voor Sterrenkunde (VVS), Oude Bleken 12, 2400 Mol, Belgium}

\authorcount{affil:LewCook}{
     Center for Backyard Astrophysics Concord, 1730 Helix Ct. Concord,
     California 94518, USA}



\maketitle

\begin{abstract}
   We observed the deeply eclipsing SU UMa-type dwarf nova
V4140 Sgr and established the very short supercycle
of 69.7(3)~d.  There were several short outbursts
between superoutbursts.  These values, together with
the short orbital period (0.06143~d), were similar to,
but not as extreme as, those of ER UMa-type dwarf novae.  
The object is thus the first, long sought, eclipsing
ER UMa-like object.  This ER UMa-like nature can naturally
explain the high (apparent) quiescent viscosity and unusual
temperature profile in quiescence, which were claimed
observational features against the thermal-tidal instability
model.  The apparently unusual outburst behavior can be reasonably
explained by a combination of this ER UMa-like nature
and the high orbital inclination and there is no need for
introducing mass transfer bursts from its donor star.
\end{abstract}

\section{Introduction}

   Dwarf novae are a class of cataclysmic variables
which show outbursts,  SU UMa-type dwarf novae are
a subclass of dwarf novae which show
superhumps during long-lasting outbursts
called superoutbursts [for general information of CVs,
dwarf novae and SU UMa-type dwarf novae and superhumps,
see e.g. \citet{war95book}].
The origin of superhumps and superoutbursts are widely
believed as a consequence of the 3:1 resonance between
the rotation in the accretion disk and the secondary star
(\cite{whi88tidal}; \cite{osa89suuma}; \cite{hir93SHperiod};
\cite{lub92SH}).

   This thermal-tidal instability (TTI) model by Osaki
has long been debated and challenged by an alternative theory
of dwarf nova outbursts --- the enhanced mass-transfer (EMT)
model.\footnote{
Although there is another pure thermal instability model
\citet{can10v344lyr}, details of this model has not yet
been published in a solid paper, we do not discuss it
further in this Letter.
}  The degree of challenge varies from SU UMa-type dwarf novae
in general (\cite{sma91suumamodel}; \cite{sma04EMT},
\cite{sma08zcha}); applications to WZ Sge-type dwarf novae,
which are a subtype of SU UMa-type dwarf novae
(e.g. \cite{pat02wzsge}; \cite{bua02suumamodel})
and to applications to special objects
(e.g. \cite{bap04v2051oph}; \cite{bap01eclipsemapping};
\cite{bap16v4140sgr}).  In SU UMa-type in general,
the detection of variation of frequencies of
negative superhumps, which are considered to arise of
a tilted disk, in the high precision Kepler data led to
the strongest support to the TTI model
(\cite{osa13v1504cygKepler};
see also \cite{osa14v1504cygv344lyrpaper3} for the final
answers to Smak's criticism).

   This Letter deals with the object (V4140 Sgr)
in the final category.  This object was studied by
\citet{bap16v4140sgr} by eclipse mapping and flickering
analysis.  \citet{bap16v4140sgr} concluded that the
temperature distribution in quiescence and the estimated
high viscosity parameter are in contradiction
with the prediction of the TTI model, and proposed
that the outbursts in this object are powered by
mass transfer bursts from its donor star.

\section{Observation and Analysis}

   The observations were carried out as a part of
a campaign led by the VSNET Collaboration \citep{VSNET}.
F.-J. Hambsch obtained snapshot observations (typically two
points a few minutes apart per night, between 2017 March 7 and
2017 November 26) and L. Cook obtained time-resolved
photometry on two nights (2017 May 4 and 7)
immediately following one of
superoutbursts.  Both observers used standard
de-biasing and flat fielding and extracted magnitudes
by aperture photometry.
The times of all observations are expressed in 
barycentric Julian days (BJD).
This object is renowned as a deeply eclipsing
cataclysmic object and we used the eclipse ephemeris
by \citep{bap16v4140sgr} after confirming that this
ephemeris expresses our observations well.

\section{Results and Discussion}

\begin{figure*}
  \begin{center}
    \FigureFile(160mm,100mm){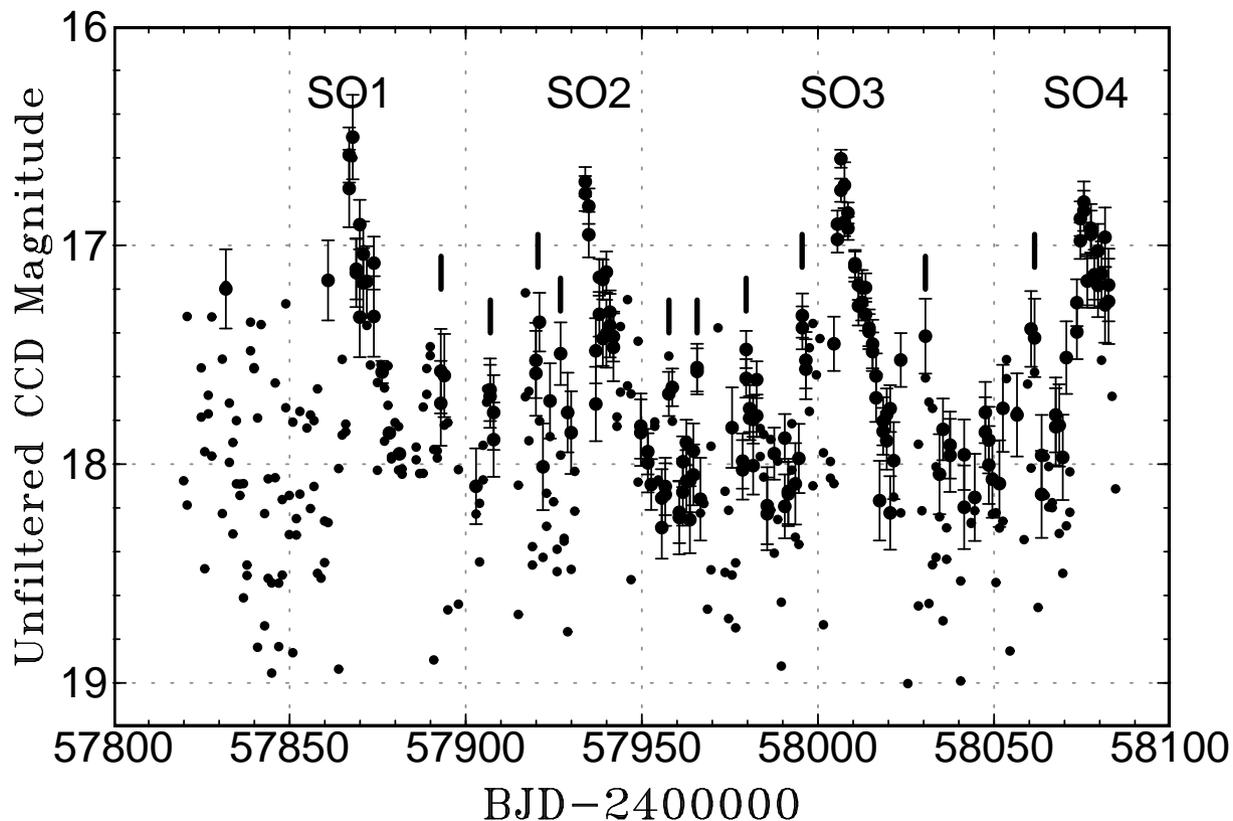}
  \end{center}
  \caption{Long-term light curve of V4140 Sgr in 2017.
     Labels SO1--SO4 represent superoutbursts and
     vertical ticks represent normal outbursts.}
  \label{fig:v4140sgrlc}
\end{figure*}

\subsection{Long-term Light Curve}

   The resultant light curve is shown in figure
\ref{fig:v4140sgrlc}.  Observations around eclipses
(orbital phases between $-$0.07 and 0.07) were removed
from this figure.
Since the object is relatively
faint for amateur instruments, individual estimates
have relatively large errors.  We selected observations
(good, reliable estimates) with errors less than 0.2 mag
as large symbols with error bars.  Other observations
(not very reliable estimates) are shown with small
filled circles without error bars to avoid complexity
of the figure.  Although errors of ``not very reliable estimates''
reached 0.9 mag in extreme cases, typical errors
were 0.3--0.5 mag when the object was below 18 mag.

   We can see from this figure that there were
very distinct four outbursts labelled with SO1--SO4.
These outbursts typically reached 16.6 mag and
linearly faded for $\sim$10~d.  The properties
of these outbursts exactly match the category of
superoutbursts (cf. \cite{war95book}).  Although
we did not perform time-resolved photometry
during these long outbursts, superhumps were almost
certainly present since superhumps were detected
during a similar long outburst in 2004 September--October
\citep{Pdot}.

   The most striking point is the shortness of
the interval (supercycle) between these superoutbursts.
A Phase Dispersion Minimization (PDM, \cite{PDM})
analysis yielded a supercycle of 69.7(3)~d.
Another striking point is the small amplitudes (1.5 mag)
of superoutbursts.

   In addition to these superoutbursts, there were
normal outbursts indicated by vertical ticks in
figure \ref{fig:v4140sgrlc}.  Normal outbursts were
difficult to detect since they are fainter than
superoutbursts and have shorter (usually 1--2~d)
durations.  We consider that not all normal
outbursts were detected.  This was partly due to
the periodic interference by the bright Moon since 
V4140 Sgr is located close to the ecliptic.
We, however, consider that the interval between
SO2 and SO3 was best observed, when many of observations
were classified as ``good, reliable estimates''.
We note the clear presence of at least four
normal outbursts (labelled with ticks) in this
interval.  The shortest intervals were 8~d (between
SO2 and SO3) and 6.4~d (between SO1 and SO2).
Considering that not all normal outbursts
could be detected, the number of normal outbursts
within a supercycle must have been relatively large
(five or more).  The gradually brightening trend
of peaks of normal outbursts between superoutbursts
is also similar to other SU UMa-type dwarf novae
(\cite{osa89suuma}; \cite{osa13v1504cygKepler}).
In \citet{Pdot9}, we reported our preliminary result
about the apparent absence of normal outbursts in V4140 Sgr.
We update the result here.

\subsection{Nature of V4140 Sgr}

   Observations indicate that V4140 Sgr has
all the characteristics of an SU UMa-type dwarf nova:
presence of superhumps during superoutbursts,
long-lasting superoutbursts occurring relatively
regularly and frequent occurrence of normal outbursts.
The only points different from ordinary SU UMa-type
dwarf novae are the shortness of the supercycle
and low outburst amplitudes.

   We know SU UMa-type dwarf novae with very short
(19--45~d) supercycles, known as ER UMa-type dwarf novae
(\cite{kat95eruma}; \cite{rob95eruma}; \cite{pat95v1159ori};
\cite{nog95v1159ori}).  Although the supercycle 69.7(3)~d
in V4140 Sgr is somewhat longer than those in
``classical'' ER UMa-type dwarf novae supercycles
in ER UMa-type dwarf novae vary both secularly and
sporadically (\cite{zem13eruma}; \cite{otu13suumacycle})
the supercycle in V4140 Sgr is only slightly longer
than the longest ``snapshot'' supercycles in classical
ER UMa-type dwarf novae [the known limit being
59~d \citet{otu13suumacycle}].
Although there exist SU UMa-type
dwarf novae with supercycles intermediate between
ER UMa-type dwarf novae and ordinary SU UMa-type dwarf novae,
these objects have much longer orbital periods
(the examples are such as SS UMi, V503 Cyg, V344 Lyr and
V1504 Cyg, see \cite{otu13suumacycle} for the references).
V4140 Sgr has a short orbital period of 0.06143~d, which
would more naturally qualify it an object analogous to
ER UMa-type dwarf novae which are known to occupy
the short-period end of the distribution of orbital periods.
Thus, we consider V4140 Sgr as the first, and long sought,
eclipsing ER UMa-like dwarf nova.  We should note, however,
that durations of superoutbursts in V4140 Sgr are shorter
than those in classical ER UMa-type dwarf novae.
V4140 Sgr probably has the properties similar to
classical ER UMa-type dwarf novae, but with a slightly
smaller mass-transfer rate.

   Regarding the low outburst amplitudes, we consider it
a combination of low outburst amplitudes in ER UMa-type
dwarf novae (typically 2--3 mag for their superoutbursts)
and the inclination effect.  In ER UMa-type dwarf novae,
the accretion disk never reaches true quiescence and
the quiescence is much brighter than in ordinary dwarf novae
(cf. \cite{osa96review}), resulting in low outburst amplitudes.
The high inclination ($i$) produces the small ($\propto \cos i$)
projected surface area of the disk and effectively
reduces the outburst amplitude.  The inclination of V4140 Sgr
is reported to be 80.2(5)$^\circ$ \citep{bor05v4140sgr}.
If the apparent brightness of the disk is indeed proportional
to $\cos i$, it should be 0.23 times that of ER UMa with
$i$=43$^\circ$.  Let's assume an extreme case that
the quiescent brightness is dominated by the hot spot\footnote{
   Although an orbital hump is not prominent in V4140 Sgr,
The light curve in \citet{bap89v4140sgr} indicated the clear
asymmetry of the eclipse, particularly the long-lasting
trailing feature of the egress phase.  This is a signature
of a geometrically broad hot spot.  We consider that
there is a significant contribution from the hot spot.
}.
This would lead to the same
apparent quiescent brightness regardless of the inclination.
An inclination effect alone leads to a reduction of outburst
amplitude 0.23 times that of ER UMa in this extreme assumption.
Since the disk is expected to significantly
contribute to the quiescent brightness of ER UMa-type
dwarf novae and it will also suffer from the inclination
effect, this ratio should be considered to be the lower
limit of the amplitude ratio.  In any case, it would be easy
to produce a low amplitude of superoutbursts (1.5 mag), which
is 0.5--0.6 times that of ER UMa.

\subsection{Implication to Outburst Mechanism}

   Such low outburst amplitudes of superoutbursts can be easily
confused with normal outbursts.  This was indeed the case
in \citet{bap16v4140sgr}, who described the ``long (80--90~d)
time interval'' which was meant to refer to the cycle length
of normal outbursts rather than the supercycle.
A similar confusion was already in the past, namely
\citet{mas01v893sco} for the eclipsing SU UMa-type dwarf nova
V893 Sco, and was corrected by \citet{kat02v893sco}.
This incorrect identification in \citet{bap16v4140sgr}
led to an assumption of long-lasting quiescent phase
and a low quiescent viscosity parameter $\alpha \sim 0.01$,
which contradicted their observation.
When we consider the ER UMa-like properties
of V4140 Sgr correctly, normal outbursts occur shortly
after the disk reaches full quiescence, and it is likely
that the disk is still turbulent following the outburst
or the hot part remaining in the (almost) quiescent disk
when \citet{bap16v4140sgr} made observations,
resulting high $\alpha$ values.

   The temperature structure is also expected to be different
from those of ordinary SU UMa-type dwarf novae, in which
the entire disk reaches the low state.  A contribution
from the hot part in the (almost) quiescent disk alters
the temperature structure and it would appear closer to
that of a steady-state ($T \propto R^{-3/4}$) disk, where
$T$ and $R$ are the temperature and the radius, respectively,
since the essence in the disk instability in ER UMa-type
dwarf novae is that the disk is almost always close to
the steady-state and only weak heating/cooling waves
travel in such a disk \citep{osa96review}.\footnote{
   Another possible interpretation of an anomalous
temperature distribution and high $\alpha$ is a result
of a disk tilt (as already proposed in \cite{Pdot9}).
It has been shown that many high-mass transfer systems
frequently show negative superhumps, particularly
in ER UMa itself (e.g. \cite{ohs14eruma}), which are
believed to arise from a tilted disk \citep{woo07negSH}.
On a tilted disk, the accretion stream hits the inner
part of the disk, resulting a high temperature
in the inner part.  Although there was no direct
observational evidence, It may have been that V4140 Sgr
spent a similar phase when observed by \citet{bap16v4140sgr}.
}
We conclude that the two important observational results
by \citet{bap16v4140sgr} for V4140 Sgr are no longer 
contradictions to the TTI model.  Rather ironically,
the same author \citep{bap89v4140sgr} earlier reported
on the same object
``The high relative brightness of the disk as compared to 
the white dwarf plus bright spot 
could be a consequence of a high accretion rate'' and
``[..] mass transfer rate greater than that due to
gravitational radiation only.  This could raise $\dot{M}$
above $\dot{M}_{\rm crit}$ and cause the disc to stay in
a high-accretion state''.  This is the very picture
(despite that $\dot{M}$ should be below, but close to,
$\dot{M}_{\rm crit}$) we reached by our new long-term
photometric observations and the presence of such
a description based on an independent set
of observations in the literature further strengthens
our interpretation.  Proper understanding of
the apparent contradiction with the claimed
low mass-transfer rate reported in
\citet{bap16v4140sgr} probably should await an accurate
parallax measurement [see e.g. the case of SS Cyg
\citep{mil13sscygdistance}].

\section*{Acknowledgements}

This work was supported by the Grant-in-Aid
``Initiative for High-Dimensional Data-Driven Science through Deepening
of Sparse Modeling'' (25120007) 
from the Ministry of Education, Culture, Sports, 
Science and Technology (MEXT) of Japan.
We are grateful to K. Isogai for processing reports
to the VSNET Collaboration.

\end{document}